\documentclass{epl} 
\newcommand{\be}{\begin{equation}}      
\newcommand{\ee}{\end{equation}}      
      
\newcommand{\bef}{\begin{figure}}      
\newcommand{\eef}{\end{figure}}      
\newcommand{\bea}{\begin{eqnarray}}    
\newcommand{\eea}{\end{eqnarray}}

\def\spose#1{\hbox to 0pt{#1\hss}}      
\def\ltapprox{\mathrel{\spose{\lower 3pt\hbox{$\mathchar"218$}}      
 \raise 2.0pt\hbox{$\mathchar"13C$}}}      
\def\gtapprox{\mathrel{\spose{\lower 3pt\hbox{$\mathchar"218$}}      
 \raise 2.0pt\hbox{$\mathchar"13E$}}}      
\def\inapprox{\mathrel{\spose{\lower 3pt\hbox{$\mathchar"218$}}      
 \raise 2.0pt\hbox{$\mathchar"232$}}}

\newcommand{\bean}{\begin{eqnarray*}}  
\newcommand{\eean}{\end{eqnarray*}}  
  
\def\lsim{\raise 0.4ex\hbox{$<$}\kern -0.8em\lower 0.62ex\hbox{$\sim$}}  
\def\gsim{\raise 0.4ex\hbox{$>$}\kern -0.7em\lower 0.62ex\hbox{$\sim$}}

\begin{document}  

\title{Universality of power law correlations in gravitational clustering}

\author{Francesco Sylos Labini\inst{1}, Thierry Baertschiger\inst{2}, 
and Michael Joyce\inst{3}}
\institute{   
\inst{1}  LPT, Universit\'e Paris-XI, B\^atiment 211, F-91405 
Orsay Cedex, France \\  
\inst{2} D\'ept.~de Physique Th\'eorique,   
            Universit\'e de Gen\`eve 24, Quai E. Ansermet,    
            CH-1211 Gen\`eve, Switzerland \\   
\inst{3} LPNHE, Universit\'e de VI, 4, Place Jussieu, Tour 33- RdC,
75252 Paris Cedec 05, France\\ 
}   
\maketitle   
   
\begin{abstract}
We present an analysis of different sets of gravitational N-body simulations, 
all describing the dynamics of discrete particles with a small 
initial velocity dispersion. They encompass very different
initial particle configurations, different numerical algorithms for the 
computation of the force, with or without the space expansion of
cosmological models. Despite these differences we find in all 
cases that the non-linear clustering which results is essentially
the same, with a well-defined simple power-law behaviour in the 
two-point correlations in the range from a few times the lower cut-off 
in the gravitational force to the scale at which fluctuations are 
of order one. 
We argue,
presenting quantitative evidence, that this apparently universal 
behaviour can be understood by the domination of the small scale 
contribution to the gravitational force, coming initially from
nearest neighbor particles. 

\end{abstract}

N-body simulations (NBS) have been widely used in cosmology to 
study  gravitational many-body dynamics in the non-linear regime
(where fluctuations in the density become large). 
The principal goal of such studies, and indeed of the theories 
of large scale structure formation, is to understand how 
from some initial conditions (IC), specified as 
a continuous density field with correlated Gaussian fluctuations, 
the large-scale structures observed in the distributions of
galaxies and clusters can arise. Specifically the problem
is to relate the very small-amplitude fluctuations in the
Universe at very early times, detected indirectly by observations
of fluctuations in the cosmic microwave background, 
to the large-amplitude fluctuations with power-law correlations 
observed today in galaxy red-shift surveys. In this letter we 
discuss the nature and origin of the power-law clustering observed
in cosmological NBS in the non-linear regime. We do this by
placing them in the context of a broader class of gravitational NBS,
identifying in these a universal behaviour in the 
non-linear clustering which develops, characterised by the
{\it exponent} of the two-point power-law correlation function.
Viewing the problem in this wider context, we argue that
the nature of clustering
in the non-linear regime has little to do with the initial 
fluctuations, or with the Universe being in expansion. Rather it
is associated with what is common to all these simulations:
their evolution in the non-linear regime is dominated by fluctuations
at small scales, which are similar in all cases at the time
this clustering develops. This corresponds to domination by
nearest neighbor interactions when the first non-linear
structures are formed.

In cosmological NBS what one wants to model is (usually) the 
evolution of the cold 
dark matter(CDM) particles  of current standard cosmological 
theories. In practice numerical limitations mean that the mass
represented by the ``particles'' in  NBS (which must simulate
a significant portion of the Universe) is typically of order 
that of a galaxy i.e. many orders of magnitude larger than the 
microscopic mass of a CDM particle. Thus a ``particle'' should
represent a collisionless fluid element rather than a physical
particle. In practice however what one simulates in almost all
cases is effectively a system of point particles: the smoothing
scale $\epsilon$ introduced to cut-off the gravitational force is much
smaller than the initial interparticle distance. At most certain
algorithms (e.g. `particle-mesh' type) make this scale
comparable to the initial interparticle distance.

In a series of papers Melott \& collaborators \cite{me90,smss98,kms96} have
discussed the effects of discretization in NBS, showing 
discrepancies in the dynamical evolution described by different
algorithms. In particular they have questioned 
the capacity of high resolution NBS (HRNBS) to describe 
correctly the evolution of a collionless self-gravitating
fluid. In a recent article
\cite{bjsl02} we have developed the central point touched on 
in this work further. Rather than considering the problem of whether,
or in what range, such codes can describe fluid-like evolution, 
we have concentrated on what is actually described by these 
HRNBS.  On the basis of an analysis 
with real space statistics of the Virgo consortium's cosmological 
HRNBS \cite{virgo}
we make the case that discreteness is not only important because it 
introduces physical effects which should be absent in the 
fluid-like growth of perturbations, but that it is 
an {\it essential} element in the formation
of power-law correlated structures.  This is one of the 
points for which we produce further evidence in this
paper, making use of our analysis of a wider class of HRNBS simulations.

\begin{table}
\label{tab1}
\caption{Details of the NBS analysed. 
(N)E indicates (no) expansion. See text for 
explanation}
\begin{center}
\begin{tabular}{ccccccc}
Simulation        & N      & L       & $\langle \Lambda_i \rangle $ 
& $\epsilon$& Code  & Ref.    \\  
\hline
POISSON   NE        &$32^3$  & $32 $       &  1         &   0.01       &  Tree code& \cite{bot02}    \\
POISSON   NE        &$8^3$   &  $8 $       &  1         &   0.01      &  Tree code&     \\
SHUF.LAT  NE        &$8^3$   &  $8 $       &  1         &   0.1       &  Tree code&      \\
SHUF.LAT  NE        &$32^3$  & $32 $       &  1         &   0.1       &  Tree code&      \\
SCDM       E        &$256^3$  &$239.5 $    &  0.94         &  0.036    &  AP$^3$M     & \cite{virgo} \\
SCDM       E        &$256^3$  & $85.4 $    &  0.33      &  0.036    &  AP$^3$M     & \cite{virgo}    \\
$\Lambda$CDM  E     &$256^3$  & $239.5$    &  0.94         &  0.036    &  AP$^3$M     & \cite{virgo}    \\
$\tau$CDM  E        &$256^3$  & $239.5$    &  0.94         &  0.036    &  AP$^3$M     & \cite{virgo}    \\
\hline
\end{tabular}
\end{center}
\end{table}

The characteristics of the NBS we analyse are summarised in Table 1.
The parameters 
characterising the simulations vary greatly: the number of 
particles N varies by almost {\it five} orders of magnitude,
while the ratio of the smoothing length in the gravitational
force $\epsilon$ to the initial mean interparticle distance
$\langle \Lambda_i \rangle $ varies by a factor of ten (and is
much smaller than unity for all of our chosen 
simulations, as discussed above). The first set are simulations
of the purely Newtonian case, without any cosmological space
expansion 
\cite{bot02,botphd}. The second set, which are NBS 
of the 
Virgo Consortium \cite{virgo} are of the evolution in specific
cosmological models, characterised by different values of
the various free parameters in currently studied models. 
These sets differ also in the algorithms which are used 
for the calculation of the gravitational force in each case,
the former using a tree code\footnote{http://www.mpa-garching.mpg.de/gadget/}
and the latter an adaptive P$^3$M code.
Note that while the lengths characterising the initial configurations 
in Table 1 are given in terms of an arbitrary length unit, in 
the cosmological sets the length $L$ corresponds to the side 
of the simulated box in $Mpc$. In this context one must choose
a physical scale for the box, as the IC fix a length scale 
(prescribing an amplitude for density fluctuations at a certain
scale at the initial time); the mass density of the Universe 
then fixes the mass of the simulated particles, and the simulation 
is run always for a time corresponding roughly to the age of 
the Universe. While in the first set the simulations are simply
run until the clustering becomes affected by the box size, in
the cosmological context the study is limited to a (small, for 
most of the NBS) part of the range 
of time evolution which could be correctly described by the
simulation.

The other important difference between these simulations is in
their IC. The first (non-cosmological) set starts with two very 
different configurations: a Poisson distribution of points and 
a `shuffled lattice' distribution. The latter is produced by applying 
a stochastic uncorrelated displacement to each point of a perfect 
cubic lattice with unitary lattice constant, the displacement vector 
being random both in orientation and length, the latter being 
sampled from a uniform probability distribution up to a maximum
displacement $|\vec{\eta}| \ll \langle \Lambda_i \rangle$. 
In the cosmological 
simulations IC are generated
in a very particular way: to represent the small initial fluctuations
(typically $\delta \rho /\rho \sim 10^{-2}$) in the CDM 
fluid correlated displacements are applied to an initially
``uniform'' distribution (either a `glassy' 
configuration \cite{virgo}, obtained by evolving first with
the sign of gravity reversed, or, more often, 
a simple lattice). The nature of the 
correlation in the displacements is determined by the power 
spectrum, which varies from model to model (the $\tau$CDM
model in Table 1, for example, differs from the SCDM only 
in this point). We note that these different IC cover a 
very wide range in terms of their power spectra $P(k)$. For the 
Poisson IC we have a constant $P(k)$, while the shuffled
lattice \cite{hz} has $P(k) \sim k^2$ for 
$2\pi/L < k < 2\pi/\langle \Lambda_i \rangle$.
The cosmological IC on the other hand have, in this
range, $P(k) \sim k^n$ with $-1< n < -3$.  Note also
that all these IC, except Poisson, also have
non-trivial higher order correlation properties.
We will return to this point below.

\begin{figure}
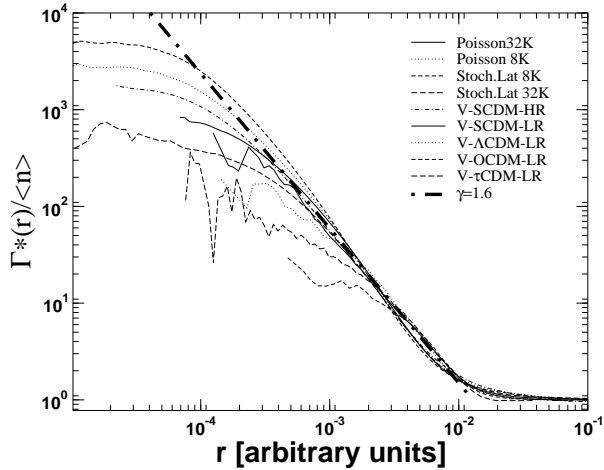

\onefigure[scale=0.35,angle=0]{FIG1.eps} 
\caption{Two-point conditional
density for the different simulations, computed using 
all particles as centers and periodic boundary conditions.)} 
\label{fig1}  
\end{figure} 

Let us now turn to our results. To characterise the clustering 
observed in the simulations we consider simply the behaviour 
of the conditional density \cite{slmp98} defined as
$\Gamma(r) = \langle n(r) n(0) \rangle / \langle n \rangle$, 
where $n(r)$ is the 
microscopic number density.
We have then considered its volume average $\Gamma^*(r)$ \cite{slmp98},
which represents the mean density of points in a ball of radius $r$
about an occupied point.  The regime of small fluctuations corresponds
in terms of it to $\Gamma^*(r) \approx  \langle n \rangle$ , where  
$\langle n \rangle= N/L^3$, the 
mean particle density in the simulation box. Within each simulation we 
observe the same qualitative behaviour of this quantity observed by 
\cite{bot02,botphd} in the 
Poisson NE simulation, and by ourselves \cite{bjsl02}
in the Virgo simulations.  Strong clustering (corresponding to 
$\Gamma^*(r) >  \langle n \rangle$) develops 
first at scales well below 
the initial mean 
interparticle separation $\langle \Lambda_i \rangle$, with a characteristic 
power-law form from a little above the smoothing scale $\epsilon$.
Subsequently the evolution is observed to be self-similar,
the same power-law form simply translating to larger and larger
scales, until (in the case that the evolution is continued that
far) boundary effects become important. In Figure \ref{fig1} we 
show in a single plot the $\Gamma^*(r)$ for each of the 
simulations in Table 1. For this comparison we have
normalised in each case $\Gamma^*(r)$ to $\langle n \rangle$ (so that
$\Gamma^*(r)/\langle n \rangle \rightarrow 1$ at large distances) 
and  
we have performed an {\it arbitrary normalization on the x-axis},
as here we are interested in the {\it slopes} of the power-law decay
and not in the range of scale where clustering develops. 
A simple power-law fits $\Gamma^*(r) \sim r^{-\gamma}$ the data  
in each case, in the range between a few times the smoothing 
scale $\epsilon$ up to a scale around 
$\Gamma^* \approx 2 \langle n \rangle $, corresponding to 
a range of scales between one and two decades.
Performing fits in this range we estimate $\gamma =1.6\pm 0.1$ for
the cosmological NBS, $\gamma=1.5 \pm 0.2$ and $\gamma=1.4 \pm 0.2$
respectively for the Poisson and shuffled lattice NBS. The
larger estimated error bars in the latter reflect the more
limited range of non-linearity in these much smaller NBS.

From our results we now argue for three conclusions
about the nature of clustering in the non-linear 
regime observed in these NBS. 
With respect to cosmological NBS, we conclude that the 
exponent characterising the non-linear clustering observed 
has essentially nothing to do 
with (i) the expansion of the Universe, or (ii) the nature 
of the small initial fluctuations imposed in the IC. 
We further present evidence for the qualitative
description of the dynamics driving this clustering 
given in \cite{bot02,botphd} based on the Poisson case, and 
in \cite{bjsl02} based on a similar analysis of the Virgo 
simulations: (iii) The non-linear clustering develops 
from the large fluctuations intrinsic to the particle 
distribution at small scales (specifically around the 
smallest resolved scale $\epsilon$). In particular we 
show here that the exponent characterising it can be 
seen to emerge at early times in the simulations 
when the evolution is well approximated as being due only 
to the interactions between nearest neighbour (NN) particles.  


The first point can be understood easily. 
Given the (observed) self-similarity in the evolution
in the non-expanding case, it is natural that it survives, 
essentially unmodified, in the cosmological case: The expansion
of the Universe, for which the characteristic time is long
compared to that of the non-linear clustering, is simply an
adiabatic rescaling of the physical scales. This can 
modify the amplitude of the correlation
function, but should not change the exponent if the latter is
indeed determined by the fluctuations at small scales.


Before discussing the second and third point it is useful
to describe in a little detail the nature of the fluctuations
in the IC of these different simulations. The initial configuration 
of particles may be characterised by its correlation functions. The 
reduced two-point correlation function, defined as 
$\tilde \xi(r) = \langle n(r) n(0) \rangle / \langle n \rangle^2 -1$,
for any point distribution can be written as \cite{hz}:
\be
\tilde \xi(r) = \delta(r)/\langle n \rangle + \xi(r) \;.
\ee
The variance in the number of points $N$
in a volume $V$ is given by 
\be
\Sigma^2 = \langle (\Delta N)^2 \rangle
=\langle n \rangle V + \langle n \rangle^2 \int_{V} d^3r_1\int_{V} d^3r_2  
\xi(|\vec{r}_1 - \vec{r}_2|) 
\label{variance1}
\ee
where $\Delta N = N-\langle N \rangle$.
The first term, which comes from the `diagonal' ($r=0$) term  
in $\tilde \xi(r)$, describes the fluctuations intrinsic to any point
distribution, while the second the fluctuations associated with 
whatever non-trivial spatial correlation there is in the distribution. 
Note that this first term is specific to {\it point} distributions,
in which fluctuations are never absent, and have large amplitude
(of order one) at small scale; in a continuous distribution, instead,
only the second term is present in both expressions
and fluctuations, which can be arbitrarily small at all scales,
are uniquely associated with correlations.

The Poisson distribution is the uncorrelated point distribution, with
$\xi=0$, in which only the first term in (\ref{variance1}) is non-zero.
Both the perfect lattice
and the `glassy' distribution used as `pre-initial' distributions (in
both the `shuffled lattice' simulation and {\it all} the cosmological
NBS) are in a specific class of discrete distributions: they are extremely
ordered distributions, with a $\xi(r)$ non-zero at all scales, describing 
delicately balanced correlations and anti-correlations, with a 
variance which grows in proportion 
to the {\it surface} of the volume \cite{hz}. 
This is in fact the slowest possible
growth of the fluctuations in any point distribution i.e. these distributions
are the most uniform possible point distributions. It is essential, however,
to note that they are {\it not} uniform. At the scales of the interparticle 
distance there are {\it large} amplitude fluctuations 
($\delta \rho/\rho \sim 1$), like in a Poisson distribution, which 
then decay rapidly ($\delta \rho/\rho \propto r^{-2}$ 
in a sphere of radius $r$, compared to $\delta \rho/\rho \propto r^{-3/2}$ 
in a Poisson distribution). 

When a displacement field is superimposed on these latter distributions
the two-point correlation properties, and fluctuations, are modified. 
Given that the displacement fields are small we can write
$\Sigma^2(r)=\Sigma^2_P(r)+ \Sigma^2_D(r)$, where the subscripts
denote the `pre-initial' and `displacement' contributions.
The new term is sub-dominant
on small scales, but can dominate at larger scales if the small amplitude
fluctuations superimposed decay less rapidly than those in the `pre-initial'
distribution. 

The universality of the non-linear exponent for which we have
produced evidence above clearly can be explained only by something
which is common to all these distributions. We will argue here that
it is essentially the first term in the variance in (\ref{variance1}),
which is common to all these distributions, which at the root
of this universality.  This interpretation is in contrast to
the standard one given of  cosmological NBS \cite{virgo}, in which
it is supposed that only the variance introduced by the
displacement field which is dynamically relevant. Thus the
evolution is understood to depend only on the small amplitude 
correlated fluctuations of the continuous CDM fluid of the 
corresponding theoretical model (but see \cite{bsl02}). 
Note, however, we are not claiming that these latter fluctuations
play {\it no} role in the evolution of the system, but only that
the exponent in the non-linear clustering does not 
significantly depend on them. 

\begin{figure}
\twofigures[scale=0.30,angle=0]{FIG2.eps}{FIG3.eps} 
\caption{Two-point conditional density for simulation from Poisson IC, one 
with the full gravitational force and one with only the contribution from
the NN.}
\label{fig2}
\caption{
$\delta F$ is the average of (the modulus of) the difference in
the force between a randomly chosen point and a second one
randomly chosen within a distance $\langle \Lambda_i \rangle$.
It is given in the figure as a function of the radius $r$ of 
the spherical region about each point from which the force
is calculated, and is normalised to its
asymptotic value $\delta F_{\rm{asym}}$. The Poisson curve
corresponds to the (non-evolved) IC in this case, while the
two other curves are for these simulations evolved to the
time at which the power-law in $\Gamma(r)$ begins to form.}
\label{fig3} 
\end{figure} 

What is important for the evolution under gravity is evidently
the relation between these intrinsic and imposed fluctuations
and the gravitational force which acts on the particles.
Let us consider first the gravitational force and the evolution
of the system in the Poisson case.  In these IC 
the gravitational force on a point is dominated by that
due to its nearest neighbour(NN) \cite{chandra}. The 
contribution from points further away cancels very efficiently 
due to statistical isotropy {\it and} the trivial higher order
correlation properties of the distribution.  To study the
role of these large NN interactions in the evolution of
clustering, we have modified an NBS to include {\it only}
this NN contribution to the gravitational force. The result
is shown in Figure \ref{fig2}. We see that the evolution of the
non-linear clustering at early times is very well described,
and in particular the exponent describing the clustering.

The other two classes of 
IC show very different qualitative 
behavior. For the shuffled lattice the (modulus of the) 
force on a particle fluctuates around 
a non-zero but small (compared to the NN contribution) value.
In the cosmological IC we have a different behaviour: the
correlation of the displacements leads to a growth of the
force as function of scale, so that the latter is in
fact dominated by the (small amplitude) perturbations 
at large scales \cite{bsl02}.   
These very different qualitative 
behaviours of the initial force distribution reflect the 
very specific correlation properties of the ``pre-initial''
lattice or glass configurations. They are configurations 
in which the force on a point is zero because there are
non-trivial correlations, in particular at small scales
(at the order of the interparticle distance) which mean 
that the forces of nearby particles exactly cancel. 
These are not, however, as we noted in \cite{bjsl02},
properties that are preserved by the gravitational evolution. 
Once particles start to move, due to the initial 
fluctuations and on a time scale determined by these, on a 
scale comparable to the interparticle separation,
we expect the NN contribution to the force to become
very important. Further, given that strong correlation 
develops starting from scales well below the initial 
interparticle separation in all these simulations, 
this contribution inevitably is important at the 
time when these structures form. Let us consider
this point more quantitatively. 
In Figure \ref{fig3} we show the average value
of the modulus of the difference in the 
(vector) forces acting on two randomly chosen
particles with a separation of less than $\langle \Lambda_i \rangle$.
It is given as a function of the radius of a sphere
centred on each particle over which we integrate 
to calculate the forces. We see that in all cases
the difference in the forces - which will determine
the relative motion of the particles - is dominated
by small scales, around the NN distance. 

This analysis applies clearly only up to the time
at which clustering develops at scales of the order
of the initial interparticle separation. At larger 
times what is observed is that the non-linear 
clustering which develops first at these
scales develops in a self-similar manner at larger
and larger scales. The self-similarity refers to
the fact that the exponent of the correlation function
remains approximately the same.
The fact that this is so suggests very strongly that 
the dynamics at play is the same as that at early times, which 
is essentially that of particles interacting by NN forces.
The evolution of the system would then be described 
as defining a  coarse-graining to new ``particles'' 
as a function of time. This is intrinsically a dynamics of a discrete
system in which the fluctuations at the smallest 
(inter-``particle'') distance are those which are
dominant in the gravitational evolution. 
Note that this does not mean that the small fluctuations 
at large scales imposed by the IC are irrelevant in the 
evolution of the system. We have noted that they will 
play a role in fixing the time scales for the evolution,
and thus in fixing the amplitude of the correlations
(and non-linearity scale) as a function of time.
This may be in keeping with what is envisaged 
in the cosmological context (through the 
linear amplification of power at large scales).
The point we have made here is  
that the fluctuations at small scales (at the
NN scale at early times) appear to be those
essential to the formation
of the power-law correlation functions.

A more quantitative description of this dynamics is
evidently needed, with the principal goal of understanding 
the specific value observed of the exponent. In the cosmological 
literature (see e.g. \cite{pee93}) the idea is widely dispersed that 
the exponents in non-linear clustering are related to that of 
the initial power-spectrum of the small fluctuations in the CDM fluid,
and even that the non-linear two-point correlation can be considered
an analytic function of the initial two-point correlations
\cite{pd96} (although, see \cite{saslaw} where more emphasis is 
put on the tendency for IC to be washed out in the 
non-linear regime).  The models used to explain the behaviour 
in the non-linear regime usually involve both the expansion 
of the Universe, and a description of the 
clustering in terms of the evolution of a continuous fluid. We 
have argued that the exponent is universal in a very wide 
sense, being common to the non-linear clustering observed 
in the non-expanding case. It would appear that the framework for 
understanding the non-linear clustering must be one in which 
discreteness (and hence intrinsically non-analytical behaviour 
of the density field) is central, and that the simple context
of non-expanding models should be sufficient to elucidate
the essential physics. 


{\bf Acknowledgements}
We thank D. Pfenniger of the Observatory of Geneva for the use of his
supercomputer Gravitor, and M. Bottaccio et al. who
kindly provided us with their simulations. We thank also
R. Durrer, A. Gabrielli, A. Melott,
M. Montuori, and  L. Pietronero for useful discussions.
FSL acknowledges the support of a Marie Curie Fellowship
HPMF-CT-2001-01443.

\end{document}